\documentclass[oneside,a4paper]{article}

\usepackage{amsthm,amsmath,amssymb,bbding,ifsym,pifont}
\usepackage{booktabs,hhline,enumerate}
\usepackage{color,subfigure}
\usepackage[pdftex]{graphicx}

\usepackage{bm,subfloat,url}

\newtheorem{example}{Example}
\newtheorem{theorem}{Theorem}
\newtheorem{lemma}{Lemma}

\newtheorem{remark}{Remark}
\newtheorem{corollary}{Corollary}
\newtheorem{definition}{Definition}

\newtheorem{algorithm}{Algorithm}

\newcommand{\R}{\mathbb{R}}
\newcommand{\C}{\mathbb{C}}
\renewcommand{\vec}[1]{\boldsymbol{#1}}

\newcommand{\ri}{\mathrm{i}}	

\renewcommand{\top}{\mathsf{T}}		
\newcommand{\cop}[1]{#1^{*}}
\newcommand{\hop}{\mathsf{H}}		

\begin{document}

\begin{center}
	{\Large Uniqueness Analysis of Non-Unitary Matrix Joint Diagonalization
	\footnote{M. Kleinsteuber and H. Shen are with the Department
	of Electrical Engineering and Information Technology, Technische 
	Universit\"at M\"unchen, M\"unchen, Germany. 
	e-mail: (see http://www.gol.ei.tum.de). \\
	\indent Authors are listed in alphabetical order due to equal contribution. \\
	} } \\[4mm]
	{\large Martin~Kleinsteuber and Hao Shen } \\[4mm]
	\today \\[9mm]
\end{center}

\begin{abstract}
	Matrix Joint Diagonalization (MJD) is a powerful approach
	for solving the Blind Source Separation (BSS) problem.
	It relies on the construction of matrices which are diagonalized by the unknown demixing matrix.
	Their joint diagonalizer serves as a correct estimate of this demixing matrix only if it is uniquely determined.
	Thus, a critical question is under what conditions a joint diagonalizer is unique. 
	In the present work we fully answer this question about the identifiability of MJD based BSS approaches
	and
	provide a general result on 
	uniqueness conditions of matrix joint diagonalization.
	It unifies all existing results 
	 which exploit the concepts of 
	\emph{non-circularity}, \emph{non-stationarity}, \emph{non-whiteness}, and 
	\emph{non-Gaussianity}.
	As a corollary, we propose a solution for complex BSS, which can be 
	formulated in a closed form in terms of an eigenvalue and a singular value 
	decomposition of two matrices.
\end{abstract}

\begin{center}
	\textbf{\small{Index Terms}} \\[2mm]
	Non-unitary joint diagonalization,
	uniqueness analysis,
	Complex Blind Source Separation (BSS), 
	Second-Order Statistics (SOS), Higher-Order Statistics (HOS).
\end{center}

\section{Introduction}
\label{sec:01}
Joint diagonalization of a set of square matrices is a
prominent algorithmic paradigm for solving the Blind Source 
Separation (BSS) problem \cite{hyva:book01}.
One critical task of the Matrix Joint Diagonalization (MJD) approach is to 
construct a set of square matrices, so that there exists a unique joint diagonalizer,
which serves as a correct demixing matrix of the BSS problem.
Although uniqueness and solvability conditions of the BSS problem are well 
known in the framework of Independent Component Analysis (ICA), cf. \cite{como:sp94},
practical identifiability of MJD based BSS approaches has not been 
systematically studied yet and is investigated thoroughly in this work.

By imposing the assumption that source signals 
are mutually statistically independent, it is well known 
that source signals can only be extracted up to arbitrary scaling and 
permutation. 
The statistical independence assumption leads to the celebrated technique of Independent 
Component Analysis (ICA) \cite{como:sp94}. 
One fundamental question is: \emph{Under what conditions on the sources
can the mixing process be uniquely identified up to permutation and scaling?}
General identifiability results of the ICA problem have been developed 
based on either the Darmois-Skitovitch theorem, cf.
\cite{como:sp94,thei:sp04}, or diagonalization of the Hessian
of the characteristic function, cf. \cite{thei:nc04}.
They provide a theoretic ground for developing ICA algorithms that
minimize the so-called \emph{contrast functions}, cf. \cite{hyva:tnn99,shen:tnn08,shen:nla11}.
Many popular ICA contrast functions originate from information theory, 
such as the mutual information \cite{pham:tit02} or the differential entropy \cite{vrin:tit07}.
Unfortunately, performance of these contrast function based 
approaches depends significantly on a correct estimation of the 
distribution of the sources, which is often an infeasible undertaking in 
application scenarios. 
Although there have been alternative non-parametric approaches developed to cope with this difficulty, cf. \cite{bosc:tnn04,shen:tsp09},
these methods go along with a high computational burden.

One simple approach to overcome the aforementioned difficulty 
is to utilize some additional properties of the sources for the separation task. 
Although the standard ICA model does not make 
any assumption on the temporal structure of the sources, 
temporal information is richly available in many real applications, 
and has been extensively exploited in developing efficient ICA algorithms, cf.
\cite{mura:nc01,hyva:tnn01}.
Specifically, these approaches utilize only selected second-order statistics (SOS) 
or higher-order statistics (HOS) of the observations, and often result in a form 
of a joint diagonalization problem of a finite number of matrices. These matrix joint
diagonalization based methods are known as the \emph{tensorial} BSS approaches.
In parallel to the general ICA theory, the present work aims to answer the 
following critical question:
\emph{Under what conditions on the matrices, which are constructed for joint 
diagonalization, can we identify the mixing process uniquely up to permutation 
and scaling?}

Early works on Matrix Joint Diagonalization in BSS are restricted 
to unitary transformations, as a whitening process on the observed 
mixtures is often used as a preprocessing step, cf. \cite{card:ieepf93}.
However, it has been shown in \cite{card:eusipco94} 
that linear BSS via a Unitary Joint Diagonaliz\-ation (UJD) 
approach may have a serious limit of degraded performance in the presence of 
additive noise. 
In particular, many criterions for joint diagonalization can be significantly 
distorted by the whitening process, cf. \cite{yere:tsp02,soul:camsap09}.
To avoid such a limit of UJD, a natural generalization 
of UJD, known as Non-Unitary Joint Diagonalization (NUJD), has
been proposed and studied with dramatically increasing attention,
cf. \cite{pham:simax01,zieh:jmlr04,soul:tsp09}.

Non-stationarity is a common property, which describes the temporal
structure of a signal. 
%
%
One simple assumption that
can be employed for BSS in this context is that
the covariance matrix of the sources varies over time.
By exploiting this property, the signals can be separated via a joint 
diagonalization of a finite set of covariance matrices within different time 
intervals \cite{pham:tsp01}.
Identifiability conditions of this approach are developed in 
\cite{card:ieepf93} for the UJD case, and in \cite{afsa:simax08} 
under a limited NUJD setting, where only the real valued ICA 
problems are consi\-dered.
Similar approaches employ also cyclo-stationarity of the sources
\cite{abed:tsp01,zeng:icassp09}, or time-frequency distributions 
at different time frequency points \cite{belo:spl04}. 


Another simple temporal concept used in BSS is the non-whiteness of sources.
Pioneering works in \cite{tong:scs90,tong:tcs91} show that real valued 
source signals with 
distinct spectral density functions are blindly identifiable by using 
only the autocorrelation of the observations.
Similar results in \cite{aiss:sp08} show that stationary colored complex
signals can be blindly separated by using a
set of autocorrelation matrices.
When source signals are both stationary and white, 
it requires more knowledge about the signals, such as higher-order 
statistics \cite{card:ieepf93}. 
In particular, third- and fourth-order cumulants have been
used and demonstrate their success in solving BSS problems, 
cf. \cite{lath:tsp01,card:icassp89,nand:spl96}.
In practice, higher order statistics 
are often rearranged in matrix form, so that
the matrix joint diagonalization approach is applicable.

All the aforementioned statistical 
properties can be used for separating both real- and complex-valued
signals.
If the sources
are {second-order non-circular} and the values of the 
 {circularity coefficients} are distinct,
 complex BSS can be 
solved effectively by a joint diagonalization of only one covariance 
matrix and one pseudo-covariance matrix, cf. \cite{lath:eusipco02}.
The corresponding method is known as Strong Uncorrelating Transform (SUT) 
\cite{erik:mlsp04}.
Unfortunately, a solution given by SUT does not in general yield a 
correct demixing of the sources in real applications, where noise 
is commonly present.
Recently, generalized SUT approaches have been proposed independently in \cite{trai:eusipco10} and \cite{shen:ica10} to jointly 
diagonalize both covariance and pseudo-covariance matrices.
%
In particular, the work in \cite{shen:ica10} demonstrates that in
the presence of noise, this generalized approach outperforms 
the state-of-the-art MJD approaches in terms of recovery quality.

To summarize,
rich literature is available in the community on developing the matrix 
joint diagonalization based BSS methods. Existing identifiability 
results are mainly focused on the SOS based approaches.
%
However, identifiability analysis for the HOS methods 
has not been addressed systematically.
In this work, we derive the uniqueness conditions of the NUJD setting.
It leads us to the most general result so far on identifiability conditions 
for the HOS based BSS methods, and an \emph{algebraic solution}, i.e. a solution that only involves Eigenvalue Decompositions (EVD) or Singular Value Decompositions (SVD). 
Furthermore, it also provides a rigorous analysis on the convergence properties of existing 
iterative algorithms \cite{shen:icassp09}.
This is due to the fact that isolated critical points of 
functions can be identified which measure the degree of joint diagonality. 
This issue is not discussed further in this paper and is subject matter of 
ongoing work of the authors.






The paper is organized as follows. 
Section~\ref{sec:02} gives a setting of the complex BSS problem and
motivates the non-unitray joint diagonalization approach as a solution to BSS. 
In Section~\ref{sec:03} we derive necessary and sufficient conditions
for the uniqueness of non-unitary joint diagonalization.
In Section~\ref{sec:04} this uniqueness result is used 
to analyze the identifiability of tensorial BSS methods and to propose
a new algebraic solution which generalizes the SUT approach and
is able to separate non-circular signals with indistinct circularity coefficients.
%

\section{Complex BSS and Matrix Joint Diagonalization}
\label{sec:02}
%
%
In this section we review the complex linear BSS problem
to make this work self-contained, together with several
second- and higher-order statistics based BSS approaches.
Thereafter, we introduce a non-unitary joint diagonalization approach
which is general enough to unify all the existing approaches in the literature.

\subsection{Notations}
\label{sec:21}
%
%
We denote by $(\cdot)^{\top}$ the matrix transpose, 
by $(\cdot)^{\hop}$ the Hermitian transpose, and by $\cop{(\cdot)}$ the (entry-wise) complex 
conjugate. Furthermore, $|z| = \sqrt{z \cop{z}}$, $\Re z$ and $\Im z$ denotes
the modulus, the real part and the
imaginary part of $z \in \mathbb{C}$, respectively. The complex unit is denoted by $\ri:=\sqrt{-1}$.
$(\cdot)^{\dag}$ stands optionally for either the matrix transpose or the 
Hermitian transpose.
Matrices are denoted with capital Roman and Greek letters, e.g. $A, \Omega$. 
Vectors are in lower case bold face, e.g. $\vec{s}, \vec{\omega}$.
The expectation value of a random variable is denoted by $\mathbb{E}[\cdot]$.

By $Gl(m)$ we denote the set of all invertible $(m \times m)$-matrices.
$I_{m}$ is the $(m \times m)$-identity matrix, and 
the sets of all unitary and real orthogonal $(m \times m)$-matrices are defined as%
\begin{align}
	& U(m) := \{ X \in Gl(m) | X^{\hop}X = I_{m} \} \quad\text{and}\\
	& O(m) := U(m) \cap \R^{m \times m},
\end{align}
respectively. The set of all complex orthogonal $(m \times m)$-matrices is given by
\begin{align}
	& O(m,\mathbb{C}) := \{ X \in Gl(m) | X^{\top}X = I_{m} \}.
\end{align}
For $C \in \mathbb{C}^{m \times m}$ and $X \in Gl(m)$, we define the linear transformations
\begin{subequations}
	\begin{align}
		C \mapsto & X C X^{\top}, \\
		C \mapsto & X C X^{\hop},
	\end{align}
\end{subequations}
as the transpose congruence transform and Hermitian congruence transform,
respectively.
Finally, we denote by $\oplus$ the exclusive disjunction operator. 

\subsection{Properties of Complex Signals}
In this work we model a complex signal $s(t)=x(t)+ \ri y(t)$ as a complex 
stochastic process indexed by the variable $t$ with real $x(t)$ and $y(t)$. 
Let $[s(t_{1}),\ldots,s(t_{n})]^{\top}$ be an $n$-dimensional 
induced random vector of the signal $s(t)$. 

\subsubsection{Stationarity}
A signal $s(t)$ is said to be \emph{completely stationary} if
the joint probability distribution 
of $[s(t_{1}),\ldots,s(t_{n})]^{\top}$ is identical to the joint probability 
distribution of $[s(t_{1}-\tau),\ldots,s(t_{n}-\tau)]^\top$ for any $n$, cf. \cite{prie:book88}.
 A real signal $x(t)$ is said to be \emph{weakly stationary}, if the following holds:
\begin{enumerate}[(i)]
	\item $\mathbb{E}[x(t)] = \mathbb{E}[x(t+\tau)]$ for all $\tau \in \R$ and
	\item $\mathbb{E}[x(t_1) x(t_2)]=\mathbb{E}[x(t_1+\tau) x(t_2+\tau)]$.
\end{enumerate}
The first property states that the mean of the signal is constant, and the second that the
correlation only depends on the time difference $t_{1}-t_{2}$.

\subsubsection{Circularity}
A complex signal $s(t)=x(t)+ \ri y(t)$ is said to be \emph{(weakly) circular},
if $s(t)$ and ${\rm e}^{\ri \alpha} s(t)$ have the same probability distribution.
The circularity assumption implies $\mathbb{E}[{s}(t)^{2}]={\rm e}^{2 \ri \alpha}\mathbb{E}[{s}(t)^{2}]$ for all $\alpha$, i.e. $\mathbb{E}[{s}(t)^{2}] = 0$.
Given a complex signal $s(t)$ with a bounded variance, i.e. 
$\mathbb{E}[|{s}(t)|^{2}] < \infty$, the following quantity
\begin{equation}
\label{eq:circ}
	\lambda_{s(t)} := \frac{|\mathbb{E}[{s}(t)^{2}]|}{\mathbb{E}[|{s}(t)|^{2}]}
\end{equation}
is referred to as the \emph{circularity coefficient} of $s(t)$.
The definition of circularity can be extended straightforwardly to the case of 
multiple signals.
Let $\vec{s}(t) \in \mathbb{C}^{m}$ be a vector consisting of $m$ signals.
Then $\vec{s}(t)$ is circular if $\vec{s}(t)$ and ${\rm e}^{\ri \alpha} \vec{s}(t)$ have the same probability distribution.

A signal $s(t)$ is said to be \emph{completely circular} if 
the induced random vector $[s(t_{1}),\ldots,s(t_{n})]^{\top}$ is circular for all $n \in \mathbb{N}$. A signal is \emph{circular of order 
$n$} if the induced vectors of order lower or equal to $n$ are circular, cf. 
\cite{ambl:sp96b}.

\subsubsection{Whiteness}
A real signal $x(t)$ is said to be \emph{white} if 
\begin{enumerate}[(i)]
	\item $\mathbb{E}[x(t)] = 0$;
	\item $\mathbb{E}[x(t_1) x(t_2)]= c \delta(t_1-t_2)$,
\end{enumerate}
where  $\delta$ is the \emph{Kronecker delta} function and $c$ some positive constant.
We refer to \cite{bond:tsp95} for generalization of the concept of whiteness to higher order.

\subsection{Complex Linear BSS Model}
\label{sec:22}
Let $\vec{s}(t) = [s_{1}(t), \ldots, s_{m}(t)]^{\top}$
be an $m$-dimensional mutually statistically independent complex signal. The noise-free instantaneous 
linear complex BSS model is given by
\begin{equation}
\label{eq:201}
	\vec{w}(t) = A \vec{s}(t),
\end{equation}
where $A \in Gl(m)$ 
is the mixing matrix and $\vec{w}(t) = [w_{1}(t), \ldots,$ 
$w_{m}(t)]^{\top}$ presents $m$ observed linear 
mixtures of $\vec{s}(t)$. 
Without loss of generality, we assume that the sources 
$\vec{s}(t)$ have zero mean, i.e.
$\mathbb{E}[\vec{s}(t)] = 0$, cf. \cite{hyva:book01}.

The task of the linear complex BSS problem \eqref{eq:201} 
is to recover the source signals $\vec{s}(t)$ by estimating 
the mixing matrix $A$ or its inverse $A^{-1}$
only based on the observations $\vec{w}(t)$ via the demixing model 
\begin{equation}
\label{eq:202}
	\vec{y}(t) = X^{\hop} \vec{w}(t),
\end{equation}
where $X^{\hop} \in Gl(m)$ is the demixing 
matrix, which is an estimation of $A^{-1}$, and $\vec{y}(t)$ represents the corresponding extracted 
signals.
The statistical independence assumption provides various
statistical properties of sources to identify the demixing
matrix. 
The widely used properties include \emph{non-circularity}, 
\emph{non-stationarity}, \emph{non-whiteness}, and \emph{non-Gaussianity}.

\subsection{Second-Order Statistics Based ICA Approaches}
\label{sec:23}
%
%
In this subsection, we briefly review the second-order statistics 
based ICA approaches and motivate our general
approach of joint diagonalization.

Given the mixing model \eqref{eq:201}, the covariance matrix of the 
observations $\vec{w}(t)$ is computed as
\begin{equation}
\label{eq:115}
		C_{\vec{w}}(t) := \mathbb{E}[\vec{w}(t) 
		\vec{w}^{\hop}(t)]
		= A \underbrace{\mathbb{E}[\vec{s}(t) 
		\vec{s}^{\hop}(t)]}_{=:C_{\vec{s}}(t)}A^{\hop}
\end{equation}
where the covariance matrix of the sources $C_{\vec{s}}(t)$ is dia\-gonal 
and non-negative following the statistical independence assumption.
When the source signals are assumed to be non-stationary, i.e. 
$C_{\vec{w}}(t_{i}) \neq C_{\vec{w}}(t_{j})$ for $t_{i} \neq t_{j}$,
the demixing matrix is expected to be identifiable via a joint diagonalization of
a set of covariance matrices at different times.

In order to separate stationary but non-white signals, one possibility is to use the
non-zero autocorrelations at different time instances $t_{1}$ and $t_{2}$ with 
$t_{1} \neq t_{2}$, namely
\begin{equation}
	\widetilde{C}_{\vec{w}}(t_{1},t_{2}) := \mathbb{E}[\vec{w}(t_{1}) 
	\vec{w}^{\hop}(t_{2})] 
	= A \widetilde{C}_{\vec{s}}(t_{1},t_{2}) A^{\hop}.
\end{equation}
Note that, although the autocorrelation matrix $\widetilde{C}_{\vec{s}}(t_{1},t_{2})$ 
of the sources is 
still diagonal, it is not real in general.
In other words, the autocorrelation matrix of the observations
is generally \emph{not} a Hermitian matrix and consequently \emph{not} positive definite either.
Similarly as above, the demixing matrix is expected to be identifiable
via a joint diagonalization of a set of autocorrelation 
matrices with different time pairs.

If the signals have a non-trivial imaginary part, additional properties can be employed for BSS.
Besides the standard covariance matrix \eqref{eq:115},
a similar statistical quantity of complex valued signals, known as
\emph{pseudo-covariance matrix}, is
defined as 
\begin{equation}
	R_{\vec{w}}(t) := \mathbb{E}[\vec{w}(t) 
	\vec{w}^{\top}(t] =
	A R_{\vec{s}}(t) A^{\top}.
\end{equation}
The works in \cite{lath:eusipco02,erik:mlsp04} have shown that,
if the sources are all non-circular with distinct 
circularity coefficients \eqref{eq:circ}, i.e. distinct diagonal entries of $R_{\vec{s}}(t)$, 
the demixing matrix can be successfully 
identified by jointly diagonalizing both the covariance and the
pseudo-covariance matrix.
The resulting algebraic solution, referred to as \emph{Strong Uncorrelating Transform}, 
provides a simple answer to the complex BSS problem.
However, it fails in separating non-circular signals with same circularity 
coefficients. To overcome this problem, one can either utilize iterative contrast function 
based algorithms or employ some additional information, as for example
the \emph{pseudo-autocorrelation matrix} of the signals, which is defined
as
\begin{equation}
	\widetilde{R}_{\vec{w}}(t_{1},t_{2}) := \mathbb{E}[\vec{w}(t_{1}) 
	\vec{w}^{\top}(t_{2})] = 
	A \widetilde{R}_{\vec{s}}(t_{1},t_{2}) A^{\top}.
\end{equation}
Note that both the pseudo-covariance and pseudo-auto\-correlation matrix
are complex symmetric.
Recent work in \cite{lixi:tsp11} considers the problem of jointly 
diagonalizing a set of both auto-correlation and pseudo-autocorrelation 
matrices.
The identifiability results for this approach are still lacking in the literature and follow from our main result in Section \ref{sec:03}. 

%
%
\subsection{Higher-Order Statistics (Tensor) Based ICA Approaches}
\label{sec:24}
In many real applications, second-order statistics may not be sufficient 
to accomplish the task of separation.
In these situa\-tions higher-order statistics can be exploited.
For example, statistically independent non-Gaussian signals
can be blindly separated by using the fourth-order
\cite{card:ieepf93}, or higher-order cumulants, cf. 
\cite{como:book10,lixi:sp11}. 

Recalling the model as given in \eqref{eq:201}, the $k$-th order cumulant
tensor of the sources $\vec{s}(t)$, denoted by $\mathcal{C}_{\vec{s},
\vec{\iota}}^{(k)} \in (\mathbb{C}^{m})^k$,
is defined with its $(i_{1},\ldots,i_{k})$-th entry by
\begin{equation}
\label{eq:cumu}
	\begin{split}
		& (\mathcal{C}_{\vec{s},\vec{\iota}}^{(k)})_{i_{1}\ldots i_{k}} 
		:= \operatorname{cum}\left(s_{i_{1}}^{(*)}\!(t),
		\ldots,s_{i_{k}}^{(*)}\!(t)\right) \\
		= & \!\sum\limits_{p=1}^{k}(-1)^{\!^{p-1}}\!(p\!-\!1)!\,
		\mathbb{E}\Big[\!\!\!\prod\limits_{\;q \in J_{1}}\!\!\!s_{q}^{(*)}\!(t)\Big]
		\!\!\cdot\ldots\cdot\!
		\mathbb{E}\Big[\!\!\!\prod\limits_{\;q \in J_{p}}\!\!\!s_{q}^{(*)}\!(t)\Big],
		\!\!
	\end{split}
\end{equation}
where $\vec{\iota} = [\iota_{1},\ldots,\iota_{k}] \in \{0,1\}^{k}$ is a binary vector which enables or disables  complex conjugate
in each dimension, i.e.
\begin{equation}
	\iota_{i} = \left\{\!
	\begin{array}{ll}
		0 &\quad \text{no~complex~conjugate}, \\
		1 &\quad \text{complex~conjugate}.
	\end{array}
	\right.
\end{equation}
%
%
The summation in \eqref{eq:cumu} involves all possible partitions 
$\{J_{1},\ldots,J_{p}\}$ ($1 \le p \le k$) of the indices $\{i_{1},\ldots,i_{k}\}$.
%
We refer to \cite{como:book10,mend:pieee91} for further details regarding
higher-order cumulant tensors.

Now, by varying two selected indices, say $(i_{p},i_{q})$, with all other 
indices fixed, we obtain one \emph{cumulant matrix} or \emph{cumulant slice}
of $\mathcal{C}_{\vec{s},\vec{\iota}}^{(k)}$,
denoted by $(\mathcal{C}_{\vec{s},\vec{\iota}}^{(k)})_{\{p,q\}} \in 
\mathbb{C}^{m \times m}$. 
The assumption that sources are mutually statistically 
independent implies that all off-diagonal entries of the cumulant tensors of 
any order must be zero, i.e., the cumulant matrices 
$(\mathcal{C}_{\vec{s},\vec{\iota}}^{(k)})_{\{p,q\}}$ are diagonal for all $p\neq q$.
Multilinear properties of the cumulant tensors lead to
\begin{equation}\label{eq:cumulus}
	(\mathcal{C}_{\vec{w},\vec{\iota}}^{(k)})_{\{p,q\}} = A 
	(\mathcal{C}_{\vec{s},\vec{\iota}}^{(k)})_{\{p,q\}} A^{\dag},
\end{equation}
where $(\cdot)^{\dag}$ is determined by the construction \eqref{eq:cumu}.
Similarly, by exploiting higher-order non-stationarity or 
higher-order circularity of the sources, the ICA problem is 
formulated as jointly diagonalizing a set of slices of the 
cumulant tensors via either Hermitian congruence or transpose 
congruence.
Note, that up to date
only specific cumulant tensors have been considered for BSS \cite{lixi:sp11}
via joint diagonalization, that end up with $(\cdot)^{\dag}$ being Hermitian conjugate
in Equation \eqref{eq:cumulus}.

In this work, we answer the question 
 on the identifiability of the BSS problem
based on the joint diagonalization of a finite set of higher-order statistics
matrices.


\subsection{A Unified NUJD Approach}

We summarize the above observations in a unified approach for non-unitary
joint diagonalization.
Let $\{C_{i}\}_{i=1}^{n}$ be a set of $m{\times}m$ complex matrices, 
constructed by 
\begin{equation}
\label{eq:205}
	C_{i} = A \Omega_{i} A^{\dag_{i}}, \quad i=1, \dots n,
\end{equation}
where $\Omega_{i} = \operatorname{diag}\big(\omega_{i1}, 
\ldots, \omega_{im} \big) \in \mathbb{C}^{m \times m}$ and
$\Omega_{i} \neq 0$.
Note, that Equation \eqref{eq:205} allows mixtures of both Hermitian congruence and transpose congruence transformations.
The task is to find a matrix 
$X \in Gl(m)$ such that the matrices
\begin{equation}
\label{eq:206}
	\left\{ \left. X^{\hop} C_{i} ( X^{\hop} )^{\dag_{i}} 
	\right| i = 1, \ldots, n \right\}
\end{equation}
are simultaneously diagonalized. 
%
Our present work concentrates on developing the uniqueness conditions
on the set of $\Omega_{i}$'s, such that the matrix $A$ is  
identifiable up to permutation and diagonal scaling.

Note that in the Hermitian congruence case, $X^\hop C_{i} X$, $i=1,\dots,n$, are 
diagonal if and only if $X$ 
simultaneously diagonalizes the Hermitian and the skew-Hermitian 
part of the $C_{i}$. 
Namely, by considering the real and the imaginary part
of $\Omega_i$ in Equation~\eqref{eq:205} that corresponds to 
$C_i=A \Omega_i A^\hop$ separately, we can construct
two Hermitian matrices as
\begin{align}
	& C'_{i}=A \Re \Omega_{i} A^{\hop} 
	\qquad\text{and}\\
	& C''_{i}=A \Im \Omega_{i} A^{\hop}.
\end{align}
Therefore, without loss of generality,
we study an equivalent formulation of Problem \eqref{eq:205}
by restricting $\Omega_{i}$ to be real diagonal
whenever $(\cdot)^{\dag}$ is the Hermitian transpose.

Clearly, the mixing matrix can only be identified up to permutation and scaling. 
We define the set of all column-wise permutated diagonal $(m \times m)$-matrices by
\begin{equation}
\label{eq:ambi}
	\begin{split}
		\mathcal{G}(m):= \{DP~|~& D \in Gl(m)\text{~is~diagonal~and~}\\
		& P \text{~is~a~permutation~matrix}\}.
	\end{split}
\end{equation}
Since $\mathcal{G}(m)$ admits a matrix group structure, we 
can define the following 
equivalence class on $\mathbb{C}^{m \times m}$, cf. \cite{belo:tsp97}.
\begin{definition}
	[Essential Equivalence] Let $X, Y \in Gl(m)$, then $X$ is
	said to be \emph{essentially equivalent} to $Y$, and vice versa, if there 
	exists $E \in \mathcal{G}(m)$ such that
	\begin{equation}
		\label{eq:ee}		
			X = Y E.
	\end{equation}
	Moreover, we say that the solution of a matrix equation is \emph{essentially unique}, if the equation admits a unique solution on the set of equivalence classes.
\end{definition}
Since 
\begin{equation}
	X^{\hop} C_{i} (X^{\hop})^{\dag_{i}} = (X^{\hop} A) \Omega_{i} 
	(X^{\hop} A)^{\dag_{i}},
\end{equation}
we assume without loss of generality for further studies that the 
$C_{i} = \Omega_{i}$, $i=1,\dots,n$, are already diagonal.
Thus, the identifiability analysis is restricted to investigating under what conditions
the unit equivalence class $\mathcal{G}(m)$ admits the only solutions
to the simultaneous diagonalization problem \eqref{eq:206}.
%
\section{Uniqueness of Non-Unitary Joint Diagonalzition}
\label{sec:03}
In this section we present the main results on the uniqueness analysis of 
the NUJD problem given by Equations~\eqref{eq:205} and \eqref{eq:206}. 
In contrast to existing results on joint diagonalization, we do not 
assume the matrices to be real as in \cite{afsa:simax08}, 
positive definite as in \cite{pham:simax01}, nor do we restrict 
the number of matrices to two as in \cite{parr:jmlr03,olli:sp09}. 
For the sake of readability, we outsource the proofs of the results 
to the appendix. 

The identifiability results require a notion of collinerarity for diagonal matrices.
Let $Z_{i}$, $i = 1, \ldots, n$, denote $n$ complex diagonal $(m \times m)$-matrices with
the diagonal entries $z_{i1}, \ldots, z_{im}$. For a fixed diagonal 
position $k$, we denote by $\vec{z}_{k} := [z_{1k}, \ldots, z_{nk}]^{\top}
\in \mathbb{C}^{n}$ the vector consisting of the $k$-th diagonal element 
of each matrix, respectively.
Recall that the cosine of the complex angle between two vectors 
$\vec{v},\vec{w} \in \C^n$ is computed as 
\begin{equation} 
\label{eq:complexcosine}
	c (\vec{v},\vec{w}):= \left\{ \begin{array}{cl} \frac{\vec{v}^\hop \vec{w}}{\|\vec{v} \| \|\vec{w}\|} & \text{ if }
	\vec{v}\neq 0 \land  \vec{w} \neq 0, \\
	1 & \text{ otherwise.}
	  \end{array} \right.
\end{equation}
where $\|\vec{v}\|$ denotes the Euclidean norm of a vector $\vec{v}$.
We
measure the collinearity of a set of $n$ complex diagonal $(m \times m)$-matrices
by means of the complex angle of the vectors
formed by stacking the entries at corresponding positions together.
Precisely, the collinearity measure for the set of $Z_{i}$'s is defined by
\begin{equation}
	\rho (Z_1, \ldots, Z_n) := \max_{1 \leq k < l \leq m} |c(\vec{z}_k,\vec{z}_l)|.
\end{equation}
Note, that $0 \leq \rho \leq 1$ and that $\rho = 1$ if and only if there exists a
complex scalar $\omega$ and a pair $\vec{z}_k, \vec{z}_l$, $k \neq l$ so that
$\vec{z}_k = \omega \vec{z}_l$.

Our first result deals with the simple situations where 
either only purely Hermitian or purely complex symmetric 
matrices are involved. 
The techniques used for deriving the uniqueness conditions for
both cases
are adapted from the work in 
\cite{afsa:simax08}, where only real symmetric matrices are considered.

\begin{theorem}
\label{thm:01}
	(a) Let $\Omega_{i} \in \mathbb{C}^{m \times m}$, for $i = 1, \ldots, n$,
	be diagonal, and let $X \in Gl(m)$ so that $X^{\hop} \Omega_{i} \cop{X}$ is diagonal as well.
	Then $X$ is essentially unique if and only if 
	$\rho(\Omega_1, \ldots, \Omega_{n}) < 1$.
%

	(b) Let $\Omega_{i} \in \mathbb{R}^{m \times m}$, for $i = 1, \ldots, n$,
	be diagonal, and let $X \in Gl(m)$ so that $X^{\hop} \Omega_{i} X$ is diagonal as well.
	Then $X$ is essentially unique if and only if 
	$\rho(\Omega_1, \ldots, \Omega_{n}) < 1$.
\end{theorem}

The above theorem answers the identifiability question of complex BSS by means of matrix joint diagonalization approaches, when either purely complex symmetric, or purely 
 Hermitian matrices are involved.
For the situations with a mixture of Hermitian and complex symmetric matrices, we continue by firstly considering 
the case of simultaneously diagonalizing \emph{one} complex symmetric and 
\emph{one} Hermitian matrix.
The following theorem generalizes a result in \cite{erik:tit06} and \cite{bene:laa84}, wherein the authors require positive definiteness of
the Hermitian matrix.

\begin{theorem} 
\label{thm:simult_diag_TWO}
	Let $X \in  Gl(m)$ and let $\Omega_{1}$ be a complex and $\Omega_{2}$ 
	a real diagonal matrix such that
	\begin{equation}
	\label{eq:thm3}
		X^{\hop} \Omega_{1} \cop{X} \text{ and } X^{\hop} \Omega_{2} X \text{ are diagonal.}
	\end{equation}
	Then $X$ is essentially unique if and only if %
	\begin{equation}
	\label{eq:thm2_secB}
		|\omega_{1k}| |\omega_{2l}| \neq |\omega_{1l}| |\omega_{2k}|,
	\end{equation}
	holds for all pairs $(k,l)$ with $k\neq l$.
\end{theorem}

Finally, by considering a mixture of multiple Hermitian and complex symmetric 
matrices, the following theorem completes our answer to the uniqueness analysis 
to the NUJD problem.
\begin{theorem}
\label{thm:simult_diagpart2} 
	Let $C_{i}= X^{\hop} \Omega_i X^{*}$ for $i=1, \ldots,s$ and
	$C'_{j}= X^{\hop} \Omega'_{j} X$ for $j=1,\ldots, h$ be diagonal.
	Moreover, let
	\begin{equation}
	\label{eq:cond_thm3}
		\rho(\Omega_{1}, \ldots ,\Omega_{s})=\rho(\Omega'_{1}, 
		\ldots, \Omega'_{h})=1,
	\end{equation}
	then $X$ is essentially unique if and only if there exists \emph{no}
	pair $(k,l)$ with $k \neq l$, such that the following two conditions hold: 
	\\[-4mm]
	\begin{subequations}
	\label{eq:conditionlast}
		\begin{align}
		\text{(i)} &\quad |c(\vec{\omega}_k,\vec{\omega}_l)|=|c(\vec{\omega}'_k,\vec{\omega}'_l)|=1 \label{eq:51a} \\
		\text{(ii)} &\quad \|\vec{\omega}_k \|\|\vec{\omega}_l \|=\|\vec{\omega}'_{k} \|\|\vec{\omega}'_{l} \|. \label{eq:51b} 
		\end{align}
	\end{subequations}
	\end{theorem}

\section{Applications to Complex BSS}
\label{sec:04}
In this section, we firstly apply the uniqueness results from the previous 
section to the NUJD based complex BSS methods. %
The second application of the uniqueness results focuses on the development 
of algebraic solutions, i.e. solutions that only involve eigenvalue or singular value decompositions.
Although the algebraic approaches are in general less powerful and 
less robust to noise and estimation errors than their iterative 
counterparts, cf. \cite{shen:ica10}, 
these methods are of particular interest, as they provide simple, efficient 
solutions based on various powerful eigensolvers, cf. 
\cite{parr:jmlr03,yere:tsp12,klei:laa09}.

\subsection{Identifiability of Complex BSS}
From the main results developed in Section~\ref{sec:03},
any existing identifiability result of complex BSS follows 
straightforwardly.
However, to the best of the authors' knowledge there are no
general results, which unify HOS based NUJD approaches. \\ \indent
Let $\vec{t} := [t_{1}, \ldots, t_{m}]^{\top} \in 
\mathbb{R}^{m}$ be a set of time instances for each observed signal
$w_{i}(t)$, we define the $k$-th order auto-cumulant tensor 
of the observations $\vec{w}(t)$, cf. \cite{mend:pieee91}, 
denoted by $\mathcal{C}_{\vec{w},\vec{\iota}}^{(k)}(\vec{t})$,
with its $(i_{1},\ldots,i_{k})$-th entry
\begin{equation}
	\left(\mathcal{C}_{\vec{w},\vec{\iota}}^{(k)}(\vec{t})
	\right)_{\!i_{1}\ldots i_{k}} \!\!:= 
	\operatorname{cum}(w_{i_{1}}^{(*)}(t_{i_{1}})
	\cdot \ldots \cdot w_{i_{k}}^{(*)}(t_{i_{k}})),
\end{equation}
%
%
%
Similarly as in Equation~\eqref{eq:cumulus}, the $(p,q)$-th slice of the 
$k$-th auto-cumulant tensor with
a set of given time $\vec{t}$ is computed as
\begin{equation}
	\left(\mathcal{C}_{\vec{w},\vec{\iota}}^{(k)}(\vec{t})\right)_{\!\{p,q\}} 
	:= 
	A \left(\mathcal{C}_{\vec{s},\vec{\iota}}^{(k)}(\vec{t})\right)_{\!\{p,q\}}  A^{\dag}.
\end{equation}
The identifiability of the complex BSS problem via jointly diago\-nalizing
a set of higher-order cumulant matrices is summarized as follows.
\begin{theorem}[Identifiability of Complex ICA]
	Given the com\-plex linear BSS model as in \eqref{eq:201} and
	a set of time instances $\vec{t}_{i} := [t_{i1}, \ldots, 
	t_{im}]^{\top}$ for $i=1,\ldots,T$, then the joint 
	diagonalizer of the set
	\begin{equation}
		\left\{ \left(\mathcal{C}_{\vec{w},\vec{\iota}}^{(k)}
		(\vec{t_{i}})\right)_{\!\{p,q\}} \right\}_{
		\substack{i = 1,\ldots,T \\
		k = 2,\ldots,K \\
		\!\!\!\!\!p,q = 1,\ldots,m } }
	\end{equation}
	is essentially unique and solves the BSS problem up to 
	permutation and scaling, if and only if
	the diagonal matrices
	\begin{equation}
		\left\{ \left(\mathcal{C}_{\vec{s},\vec{\iota}}^{(k)}
		(\vec{t_{i}})\right)_{\!\{p,q\}} \right\}_{
		\substack{i = 1,\ldots,T \\
		k = 2,\ldots,K \\
		\!\!\!\!\!p,q = 1,\ldots,m } }
	\end{equation}
	fulfill one of the following three conditions:
	\begin{enumerate}[(i)]
		\item $\rho(\mathcal{S})<1$, where $\mathcal{S}$ denotes the set of cumulant 
			matrices constructed via transpose congruence, i.e.
			\begin{equation}
				\mathcal{S} := \left\{\! \left(\mathcal{C}_{\vec{s},
				\vec{\iota}}^{(k)}(\vec{t_{i}})\right)_{\!\{p,q\}} \Big|
				\iota_{p} \oplus \iota_{q} = 0 \right\};
			\end{equation}
		\item $\rho(\mathcal{H})<1$, where $\mathcal{H}$ is the set of cumulant 
			matrices constructed via Hermitian congruence, i.e.
			\begin{equation}
				\mathcal{H} := \left\{\! \left(\mathcal{C}_{\vec{s},
				\vec{\iota}}^{(k)}(\vec{t_{i}})\right)_{\!\{p,q\}} \Big|
				\iota_{p} \oplus \iota_{q} = 1 \right\};
			\end{equation}
		\item When both the previous two conditions are violated, 
			Equation~\eqref{eq:conditionlast} still holds.
	\end{enumerate}
	%
\end{theorem}

\begin{example}[Fourth-Order Cumulants]
	Recall the complex BSS model \eqref{eq:201}, the fourth-order cumulant 
	of a subset of chosen sources $(s_{i_{1}},s_{i_{2}},s_{i_{3}},s_{i_{4}})$ is computed
	explicitly as
	\begin{equation}
	\label{eq:cum4}
		\begin{split}
			\operatorname{cum}(s_{i_{1}},s_{i_{2}},s_{i_{3}},s_{i_{4}}) = 
			&\!~\mathbb{E}[s_{i_{1}}\!(t)s_{i_{2}}\!(t)s_{i_{3}}\!(t)s_{i_{4}}\!(t)] \\
			& - \mathbb{E}[s_{i_{1}}\!(t)s_{i_{2}}\!(t)] 
			\mathbb{E}[s_{i_{3}}\!(t)s_{i_{4}}\!(t)] \\
			& - \mathbb{E}[s_{i_{1}}\!(t)s_{i_{3}}\!(t)] 
			\mathbb{E}[s_{i_{2}}\!(t)s_{i_{4}}\!(t)] \\
			& - \mathbb{E}[s_{i_{1}}\!(t)s_{i_{4}}\!(t)] 
			\mathbb{E}[s_{i_{2}}\!(t)s_{i_{3}}\!(t)].
		\end{split}
	\end{equation}
	By taking into account all possible combinations of complex conjugate on each
	component, we have three different fourth-order cumulant tensors 
	\begin{subequations}
	\label{eq:cute}
		\begin{align}
			(\mathcal{C}_{\vec{w},\vec{\iota}_{1}}^{(4)})_{i_{1}\ldots i_{4}} := 
			& \operatorname{cum}(w_{i_{1}},w_{i_{2}},w_{i_{3}},w_{i_{4}}), \\
			(\mathcal{C}_{\vec{w},\vec{\iota}_{2}}^{(4)})_{i_{1}\ldots i_{4}} := 
			& \operatorname{cum}(w_{i_{1}}^{*},w_{i_{2}},w_{i_{3}},w_{i_{4}}), \\
			(\mathcal{C}_{\vec{w},\vec{\iota}_{3}}^{(4)})_{i_{1}\ldots i_{4}} := 
			& \operatorname{cum}(w_{i_{1}}^{*},w_{i_{2}}^{*},w_{i_{3}},w_{i_{4}})
			\label{eq:cum43}.
		\end{align}
	\end{subequations}
	Current works in the BSS literature only focus on the cases, where source
	signals are assumed to be \emph{harmonic}, i.e. the quantity \eqref{eq:cum43} 
	does not vanish, while the other two are equal to zero, cf. \cite{mend:pieee91}. 
	Theorem~3 in \cite{tong:tcs91} presents a result on the identifiability
	of separating harmonic sources using only the $4$-th order cumulants \eqref{eq:cum43}.
	Certainly, when the sources are \emph{non-harmonic}, i.e. all possible 
	fourth-order cumulants \eqref{eq:cute} do not vanish, 
	then the BSS problem can be still solvable via a joint 
	diagonalization of fourth-order cumulant matrices, even though the conditions 
	given in \cite{tong:tcs91} are violated. 
\end{example}

\subsection{Algebraic Solutions to Complex BSS}
\label{sec:4b}
In this subsection, we investigate a particularly simple 
solution to the complex BSS problem. It is given in closed form in terms of an eigenvalue and a singular value decomposition of two matrices. We refer to such solutions as \emph{algebraic solutions}.
These methods are of high interest, since the existence of fast eigensolvers 
turns them into very fast solvers for BSS. 
Algebraic solutions exist if either both matrices are complex symmetric or Hermitian,
and at least one is invertible, cf. \cite{parr:jmlr03}.
For the mixed case, the strong uncorrelating transform (SUT), where a combination 
of an eigenvalue decomposition and a Takagi factorization is used, provides an 
algebraic solution only if the Hermitian matrix is positive definite.
In this subsection, we extend this approach and investigate the situation of separating 
non-circular signals with non-distinct circularity coefficients, cf. \cite{shen:ica12b}.

\begin{lemma}
\label{lem:put}
Let $C_{1}, C_{2} \in Gl(m)$ be one complex 
symmetric and one Hermitian matrix, respectively, 
constructed by 
\begin{align}
\label{eq:put1}
	& C_{1} := A\Omega_{1}A^{\top}
	\qquad\text{and}\\
\label{eq:put2}&	C_{2} := A\Omega_{2}A^{\hop},
\end{align}
where $A \in  Gl(m)$, $\Omega_{1}$ is complex diagonal, and $\Omega_{2}$ 
is real diagonal.
Let $C_{2} = U \Sigma U^{\top}$ be the Takagi factorization of $C_{2}$.
	Then, \vspace{-1mm}
	\begin{enumerate}[(i)]
		\item the matrix $\widetilde{C}_{1} := \Sigma^{-1/2} U^{\hop} C_{1} U \Sigma^{-1/2}$ 
			admits a matrix factorization of the form $\widetilde{C}_{1} = 
			V \Lambda V^{\hop}$, where $V \in O(m)$ and $\Lambda$ is diagonal; 
		\item the matrix $X := U \Sigma^{-1/2} \cop{V}$ satisfies
		\begin{equation}
		X^{\hop} C_{2} \cop{X}=I \quad \text{and} \quad X^{\hop} C_{1} {X} \text{ is diagonal.}
		\end{equation}
	\end{enumerate}
\end{lemma}

As the complex symmetric matrix $C_{2}$ reflects the pseudo second-order 
statistics of complex signals, we name the matrix $X$ 
\emph{Pseudo-Uncorrelating Transform (PUT)} in referring its 
connection to SUT.
A straightforward computation shows that the matrix $V$ consists of the eigenvectors of
$\widetilde{C}_{1} \widetilde{C}_{1}^{\top}$, as \vspace{-0.2mm}
\begin{equation}
\label{eq:19}
	\widetilde{C}_{1} \widetilde{C}_{1}^{\top} 
	= V \Lambda V^{\hop} \cop{V} \Lambda V^{\top}
	= V \Lambda^{2} V^{\top}. \vspace{-0.2mm}
\end{equation}
%
Thus, if $W$ is a matrix such that $\widetilde{C}_{1} \widetilde{C}_{1}^{\top} = W 
\Lambda' W^{-1}$ and if the eigenvalues $\Lambda'$ are pairwise distinct, it 
follows by the uniqueness of the EVD, that $V=W (W^{\top} W)^{-1/2} D P$, where 
$P$ is a permutation and $D$ is diagonal with entries being $\pm 1$.
Ultimately, we summarize the procedure for computing the PUT in Algorithm \ref{algo:01}.
\begin{center}
\begin{tabular}{p{0.97\columnwidth}}
	\toprule\hline\vspace*{-2.5mm}
	\hspace*{-0.5mm} \begin{algorithm}\label{algo:01}
	Pseudo-Uncorrelating Transform (PUT)
	\end{algorithm} \\[-3.5mm]
	\hline\hline\vspace*{-1.5mm}
	\hspace*{1mm} Step 1: Construct $C_{1},C_{2}$ from the observations $\vec{w}(t)$,
		\\
	\hspace*{13mm} where $C_{1}$ and $C_{2}$ are constructed via Hermitian \\
	\hspace*{13mm} congruence and matrix
		congruence, respectively; \\[0.5mm]
	\hspace*{1mm} Step 2: Compute the Takagi factorization of $C_{2} 
		= U \Sigma U^{\top}$; \\[0.5mm]
	\hspace*{1mm} Step 3: Let 
		$\widetilde{C}_{1} := \Sigma^{-1/2} U^{\hop} C_{1} U \Sigma^{-1/2}$, 
		compute EVD of \\
	\hspace*{13mm} $\widetilde{C}_{1} 
		\widetilde{C}_{1}^{\top} = W \Lambda W^{-1}$; \\[0.5mm]
	\hspace*{1mm} Step 4: Compute $V = W (W^{\top}W)^{-1/2}$; \\[0.5mm]
	\hspace*{1mm} Step 5: Compute the PUT matrix $X = U 
	\Sigma^{-1/2} \cop{V}$; \\[1mm]
	\hline\bottomrule \\
	\end{tabular}
\end{center}

\begin{remark}
\label{rem:01}
	When the matrix $C_{1}$ is Hermitian and 
	positive definite, i.e. $C_{1}$ being the covariance matrix of the observations,
	then the entries of $\Lambda$ in \eqref{eq:19} are simply the reciprocal of the circularity
	coefficients of sources.
	Our result coincides with the identifiability condition of SUT, cf. 
	theorem~2 in \cite{erik:mlsp04}.
	\end{remark}
	\begin{remark}
	The second observation is that the SUT of an arbitrary pair of one 
	positive definite Hermitian and one complex symmetric matrix 
	does always exist, 	cf. \cite{bene:laa84}.
	In contrast, the existence of the PUT matrix is not guaranteed for an arbitrary pair of a complex symmetric and a (general) Hermitian matrix. 
	%
	%
	However, existence of SUT implies the applicability of PUT
	on an arbitrary pair of positive definite Hermitian and complex symmetric 
	matrix.
	In other words, PUT can be considered as a generalization of SUT.
\end{remark}

\begin{corollary}
\label{cor:01}
	For an arbitrary pair of one Hermitian positive definite and one non-singular
	complex symmetric matrix, a PUT matrix always exists. 
\end{corollary}


Finally, we characterize the applicability of PUT as an effective BSS technique. 
Recall the complex linear BSS model as in \eqref{eq:201}, 
let $\vec{t} := [t_{1}, \ldots, t_{m}]^{\top} \in \mathbb{R}^{m}$
represent $m$ time instances of individual observations, and denote 
by $\widetilde{C}_{\vec{s}}(\vec{t})$ and $\widetilde{R}_{\vec{s}}(\vec{t})$ 
the autocorrelation and pseudo-autocorrelation matrix of the sources $\vec{s}(t)$, respectively.
Their $(i,j)$-th entries are computed as
\begin{equation}
	\left( \widetilde{C}_{\vec{s}}(\vec{t}) \right)_{ij} := 
	\mathbb{E}[s_{i}(t_{i}) s_{j}^{*}(t_{j})],
\end{equation}
and
\begin{equation}
	\left( \widetilde{R}_{\vec{s}}(\vec{t}) \right)_{ij} := 
	\mathbb{E}[s_{i}(t_{i}) s_{j}(t_{j})].	
\end{equation}

\begin{corollary}
	%
	%
	If one of the two conditions:
	\vspace{.2cm}
	\begin{enumerate}[(i)]
		\item 
	$\Big| \Re\! \left(\! \widetilde{C}_{\vec{s}}(\vec{t}) \!\right)_{\!ii} \!\Big|
	 \Big|\! \left(\! \widetilde{R}_{\vec{s}}(\vec{t}) \!\right)_{\!jj} \!\Big| 
	\!\neq\! 
	 \Big| \Re\! \left(\! \widetilde{C}_{\vec{s}}(\vec{t}) \!\right)_{\!jj} \!\Big|
	 \Big|\! \left(\! \widetilde{R}_{\vec{s}}(\vec{t}) \!\right)_{\!ii} \!\Big|,$
	 	\item 
	$\Big| \Im\! \left(\! \widetilde{C}_{\vec{s}}(\vec{t}) \!\right)_{\!ii} \!\Big|
	 \Big|\! \left(\! \widetilde{R}_{\vec{s}}(\vec{t}) \!\right)_{\!jj} \!\Big| 
	\!\neq\! 
	 \Big| \Im\! \left(\! \widetilde{C}_{\vec{s}}(\vec{t}) \!\right)_{\!jj} \!\Big|
	 \Big|\! \left(\! \widetilde{R}_{\vec{s}}(\vec{t}) \!\right)_{\!ii} \!\Big|,$
	\end{enumerate}
	\vspace{.2cm}
	is fulfilled for all pairs $(i,j)$ with $i \neq j$,	
	then the joint diagonalizer of one Hermitian and one
	complex symmetric matrix, constructed correspondingly from 
	the observations $\vec{w}(t)$ via PUT, is essentially 
	unique and solves the BSS problem up to permutation and scaling.
\end{corollary}

\section{Conclusions}
\label{sec:05}
In this work, we study the problem of simultaneously 
diagonalizing a set of complex square matrices, and provide 
a thorough uniqueness analysis of the problem.
In particular, we focus on its application in the problem of complex linear 
BSS.
Our work not only characterizes a general result on identifiability conditions 
of the MJD based BSS methods, but also derives a generalized algebraic BSS solution, 
i.e. the PUT algorithm.
Furthermore, the present results may also have 
impact in the areas of beamforming \cite{huan:ieeesj07}, and 
direction of arrival estimation \cite{zeng:spl09}, where matrix joint 
diagonalization approaches play an important role.

\section*{Acknowledgement}
This work has partially been supported by the Cluster of Excellence
\emph{CoTeSys} - Cognition for Technical Systems, funded by the German 
research foundation (DFG).

\appendix

%


\subsection{Proof of Theorem \ref{thm:01}} \label{subsec:proof1}
%
%
(a)   First, consider the case $m=2$ and let 
   \begin{equation}
       X=\begin{bmatrix}
       x_1 & x_2 \\
       x_3 & x_4
       \end{bmatrix} \in Gl(2).
   \end{equation}
   Then 
   $X^{\hop} \Omega_{i} \cop{X}$ is diagonal for $i=1, \ldots, n$, 
   if and only if
   \begin{equation}
   \label{eq:proofeqsymm}
       \omega_{i1}^{*} x_{1} x_{2} + \omega_{i2}^{*} x_{3} x_{4}=0,
   \end{equation}
   for $i=1, \ldots, n$. 
   The corresponding system of linear equations reads as
   \begin{equation}
   \label{eq:proofeq2}
       \begin{bmatrix}
       \omega_{11} & \omega_{21} & \ldots & \omega_{n1} \\
       \omega_{12} & \omega_{22} & \ldots & \omega_{n2}
       \end{bmatrix}^{\hop}
       \begin{bmatrix} x_{1} x_{2} \\ x_{3} x_{4}
       \end{bmatrix} =0,
   \end{equation}
   which only has a unique trivial solution if and only if the 
   coefficient matrix has rank $2$. This is equivalent to 
   $\rho (\Omega_{1}, \ldots, \Omega_{n})<1$.
   The trivial solution, i.e. $x_{1} x_{2} = x_{3} x_{4} =0$,
   together with the invertibility of $X$ yields that either
   $x_{1} = 0$ and $x_{4}=0$, or, $x_{2}=0$ and $x_{3}=0$. 
   This, in turn, is equivalent to $X \in \mathcal{G}(2)$. 

   Consider now the case $m >2$.  If $\rho = 1$ then there exists 
   a pair $(k,l)$ such that $|c(\vec{\omega}_k,\vec{\omega}_l)|=1$ 
   and the same argument as above shows that $\rho = 1$ implies 
   the non-uniqueness of the joint diagonalizer.
   
   For the reverse direction of the statement, assume that the joint 
   diagonalizer $X$ is not in $\mathcal{G}(m)$. We 
   have to show that this implies $\rho = 1$.

   Now assume first that one of the $\Omega_{i}$'s, say $\Omega_{1}$, 
   is invertible. 
   Then 
   \begin{equation}
   \label{eq:proofeq4}
       X^{\hop} \Omega_{i} \cop{X} (X^{\hop} \Omega_{1} \cop{X})^{-1} 
       = X^{\hop} \Omega_{i} \Omega_{1}^{-1} (X^{\hop})^{-1}, 
   \end{equation}
   for $i=1, \ldots, n$, gives the simultaneous eigendecomposition of 
   the diagonal matrices $\Omega_i \Omega_{1}^{-1}$. Since $X \notin \mathcal{G}(m)$, 
   there exists a pair $(k,l)$ with $k \neq l$ such that 
   \begin{equation}
   \label{eq:19a}
       \frac{\omega_{ik}}{\omega_{1k}}=\frac{\omega_{il}}{\omega_{1l}}, 
   \end{equation}
   which is equivalent to $|c(\vec{\omega}_k,\vec{\omega}_l)|=1$
   and hence $\rho(\Omega_{1}, \ldots,\Omega_{n})=1$. 
   
   If all the $\Omega_{i}$'s are singular, we distinguish between \
   two cases.
   Firstly, assume that there is a position on the diagonals, say $k$,
   where all $\omega_{ik}=0$. Then $|c(\vec{\omega}_k,\vec{\omega}_l)|=1$ 
   holds true for any $k \neq l$ and thus $\rho=1$.
   Secondly, if there is no common position where all the 
   $\Omega_{i}$'s have a zero entry, there exists an invertible 
   linear combination, say $\Omega_{0}$, which can also be diagonalized via
   the same transformations.
   Then by considering a new set $\{\Omega_{i}\}_{i=0}^{n}$, the same argument 
   as from \eqref{eq:proofeq4} to \eqref{eq:19a} 
   for the invertible case applies by replacing $\Omega_1$ with $\Omega_0$.
   This completes the proof for part (a).

   %
(b)   For $m=2$, the condition that $X^{\hop} \Omega_{i} X$ is diagonal 
   for all $i=1, \ldots,n$ leads to the system of linear equations 
   \begin{equation}
   \label{eq:proofeq5}
       \begin{bmatrix}
       \omega_{11} & \omega_{21} & \ldots & \omega_{n1} \\
       \omega_{12} & \omega_{22} & \ldots & \omega_{n2}
       \end{bmatrix}^{\top}
       \begin{bmatrix} x_1 x_{3}^{*} \\ x_{2} x_{4}^{*}
       \end{bmatrix} =0,
   \end{equation}
   which admits a non-trivial solution if and only if 
   $\rho(\Omega_{1}, \ldots, \Omega_{n})$ $=1$.
   Now, $x_1 x_{3}^{*} = x_{2} x_{4}^{*}
   =0$ together with the invertibility of $X$ implies that $X$ is 
   essentially unique. 
   The case for $m > 2$ is now just as in 
   Section \ref{subsec:proof1} and is omitted here. 

\subsection{Proof of Theorem \ref{thm:simult_diag_TWO}}
   We prove an equivalent formulation of Theo\-rem \ref{thm:simult_diag_TWO}. 
   Namely, a matrix $X \in Gl(m)\setminus \mathcal{G}(m)$  
   that fulfills condition \eqref{eq:thm3} 
    exists, if and only if there exists a pair $(k,l)$ with 
   $k \neq l$ such that
   \begin{equation}
   \label{eq:thm2_equal}
       |\omega_{1k}| |\omega_{2l}| = |\omega_{1l}| |\omega_{2k}|.
   \end{equation}
   Firstly, consider the case $m=2$. From Equations \eqref{eq:proofeqsymm} 
   and \eqref{eq:proofeq5} we see that the condition \eqref{eq:thm3} 
   is equivalent to
   \begin{equation}
   \label{eq:schlonz001}
       \left\{\!\!
       \begin{array}{ll}
           \omega_{11}^{*} x_{1} x_{2} + \omega_{12}^{*} x_{3} x_{4}=&\!\!\!\!0 
           \\
           \omega_{21} x_{1} x_{2}^{*} + \omega_{22} x_{3} x_{4}^{*}=&\!\!\!\!0.
       \end{array}
       \right.
   \end{equation}
   Assume now that $X \in Gl(2)\setminus \mathcal{G}(2)$ and, 
   without loss of generality $|x_{1} x_{2}|\neq 0$.
   Then either $|x_{3} x_{4}|\neq 0$ and Equation~\eqref{eq:schlonz001} yields
   \begin{equation}
   |\omega_{11}| = |\omega_{12}| \frac{ |x_{3} x_{4}|}{|x_{1} x_{2}|} , \quad
   |\omega_{21}| = |\omega_{22}| \frac{ |x_{3} x_{4}|}{|x_{1} x_{2}|},
   %
   \end{equation}
    or $|x_{3} x_{4}|= 0$. Both cases imply Equation~\eqref{eq:thm2_equal}.

   For the other direction, let Equation \eqref{eq:thm2_equal} hold true. 
   We construct explicitly a common diagonalizer in $Gl(2)\setminus\mathcal{G}(2)$. 
   The case when either $\Omega_{1}=0$ or $\Omega_{2}=0$ is trivial and not further discussed. Equation \eqref{eq:thm2_equal} implies
   \begin{equation}
       \Omega_{1} = r \begin{bmatrix} \exp{({\rm i} \varphi_1}) & \\ & \exp{({\rm i} \varphi_2}) \end{bmatrix} \Omega_{2}, 
   \end{equation}
   with suitable $\varphi_i \in [0, 2\pi)$ and $r > 0$.
   Firstly, assume that one, and hence both, matrices $\Omega_{1}$ and 
   $\Omega_{2}$ are not invertible.
   We choose without loss of generality $\omega_{22}$ to be $0$. 
   Equation \eqref{eq:schlonz001} now implies
   $x_{1} x_{2}=0$, but $x_{3}$ and $x_{4}$ can be chosen arbitrarily. 
   Indeed, it is easily checked that in this case,
   \begin{equation}
       X:= \begin{bmatrix} 1 & 1 \\ 0 & 1 \end{bmatrix}
   \end{equation}
   is a common diagonalizer.
   
   
   Assume now that both, $\Omega_{1}$ and $\Omega_{2}$ are invertible.
   Then it is straightforwardly verified that 
   \begin{equation}
       X := \Theta \Omega_{2}^{-1/2} 
       \begin{bmatrix} 
       \exp{(-\frac{{\rm i}}{2} \varphi_1}) & 
       \\ 
       & \exp{(- \frac{{\rm i}}{2} \varphi_2}) 
       \end{bmatrix}
   \end{equation}
   is a common diagonalizer for any real orthogonal matrix 
   $\Theta \in O(2)$. 

   Now, let $m > 2$. If Equation \eqref{eq:thm2_equal} holds true, then 
   the case for $m=2$ applies and the diagonalizer is not essentially 
   unique. 

   For the reverse direction, we assume firstly that both
   $\Omega_{1}$ and $\Omega_{2}$ are not invertible. Then 
   either there exists an index pair $(k,l)$ with $k \neq l$, 
   such that Equation \eqref{eq:thm2_equal} holds true (with zeros 
   on both sides of the equation) and it follows again from the 
  case  $m=2$  that the diagonalizer is not essentially unique. 
   Or, $\Omega_{1}$ and $\Omega_{2}$ both have at most one zero 
   diagonal entry at different positions.
   This case will be treated at the end of the proof.

   Let us now consider the case where $\Omega_{2}$ is invertible.
   Assume that the diagonalizer is not essentially unique, 
   i.e. that $X$ in Equation \eqref{eq:thm3} (and hence $X^{\hop}$ and 
   $\cop{X}$) differs from a product of a diagonal and a permutation 
   matrix. 
   The uniqueness of the \emph{QR}-decomposition of the invertible matrix $X$, i.e.
   $X = Q_{X} R_{X}$, allows by further decomposing $R_{X} = D_{X} N_{X}$ with $D_{X}:={\rm ddiag}(R_{X})$ and $N_{X}:=D_{X}^{-1}R_{X}$ the unique
factorization 
   \begin{equation}
   \label{eq:iwasawa}
       X = Q_{X} D_{X} N_{X}
   \end{equation}
   with unitary $Q_{X}$, positive real diagonal $D_{X}$, and $N_{X}$ 
   being upper triangular with ${\rm ddiag}(N_{X})= I_{m}$.
   Here, $\operatorname{ddiag}(N_{X})$ forms a diagonal matrix, 
   whose diagonal entries are just those of $N_{X}$.
   %

       
   Using this decomposition, $X$ is not in $\mathcal{G}(m)$ if and only if either 
   $N_{X} \neq I_{n}$ or $Q_{X}$ is not a product of a permutation 
   matrix and a diagonal phase shift matrix. 
   By a diagonal phase shift matrix, we mean all diagonal
   matrices in $U(m)$.
   Using the invertibility assumption on $\Omega_{2}$, 
   \begin{equation}
   \begin{split}
   Z :& =
   (X^{\hop} \Omega_{2} X)^{-1} X^{\hop} \Omega_{1} \cop{X} \\ 
   & = X^{-1} \Omega_{2}^{-1}\Omega_{1} \cop{X} \\
   & = N_{X}^{-1} D_{X}^{-1} Q_{X}^{\hop} \Omega_{2}^{-1} 
   \Omega_{1} \cop{Q_{X}} \cop{D_{X}} \cop{N_{X}}
   \end{split}
   \end{equation}
   is diagonal. This yields
   \begin{equation}
   \label{eq:proofenen00}
       D_{X}^{-1} Q_{X}^{\hop} \Omega_{2}^{-1} \Omega_{1} 
       Q_{X}^{*} D_{X}^{*} = N_{X} Z (N_{X}^{*})^{-1},
   \end{equation}
   where the matrix is symmetric on the left hand side and upper 
   triangular on the right hand side. This leads us to two conclusions, 
   namely that
   \begin{equation}
       N_{X} Z (N_{X}^{*})^{-1} \quad \text{is diagonal}
   \end{equation}
   and 
   \begin{equation}
       D_{X}^{-1} Q_{X}^{\hop} \Omega_{2}^{-1} \Omega_{1} Q_{X}^{*} D_{X}^{*} \quad \text{is diagonal.}
   \end{equation}
Since $D_{X}=D_{X}^{*}$ is real and diagonal, the last Equation implies that
\begin{equation}\label{eq:takagigedoens}
  \widetilde{R}=Q_{X}^{\hop} \Omega_{2}^{-1} \Omega_{1} Q_{X}^{*} \quad \text{is diagonal}
   \end{equation}
and hence
 \begin{equation}
   \label{eq:proofenen}
       \widetilde{R} = N_{X} Z (N_{X}^{*})^{-1},
   \end{equation}

   Let us have a closer look at Equation \eqref{eq:takagigedoens}. 
   By introducing suitable diagonal phase shift matrices 
   $\Phi_1$ and $\Phi_2$ we have
   \begin{align}
   \label{eq:proof28}
       \Phi_1 Q_{X}^{\hop} \Phi_2 \!
       \begin{bmatrix} 
           | \omega_{11} / \omega_{21} | & & \\
           & \ddots & \\ 
           & & | \omega_{1m} / \omega_{2m} |  
       \end{bmatrix} \!\Phi_2 Q_{X}^{*} \Phi_1 = R
   \end{align}
   where $R$ is diagonal with real and nonnegative entries.
   Note that Equation~\eqref{eq:proof28} gives a Takagi factorization of 
   $R$.
   If $Q_{X}$ differs from a product of a permutation matrix and 
   a phase shift matrix, the uniqueness of the Takagi factorization 
   now implies that (at least) two diagonal entries have to coincide 
   and consequently Equation \eqref{eq:thm2_equal} follows.

   Assume now that $N_{X}$ differs from the identity, 
   and let its $(k,l)$-th entry, say $z$, differ from $0$.
   Now $\widetilde{R}=\Phi_1^{*2} R$ and consequently
   Equation \eqref{eq:proofenen} yields
   \begin{equation}
   \label{eq:prolaber}
       (N_{X})^{-1} \Phi_{1}^{*2} R N_{X} = Z.
   \end{equation}
   Note that, by the special structure of $N_{X}$, namely upper 
   triangular with ones on the diagonal,
   this immediately implies $Z= {\Phi}_1^{*2} R$.
   Thus, the $(k,l)$-th entry of equation \eqref{eq:prolaber} reads as
   \begin{equation}
       z ({\Phi}_1^{*2} R)_{kk}= z ({\Phi}_1^{*2} R)_{ll}. 
   \end{equation}
   Taking absolute values, this implies $|R_{kk}|=|R_{ll}|$ for the corresponding diagonal entries of $R$ and Equation \eqref{eq:thm2_equal} follows.

   Now let us get back to the case where exactly one diagonal 
   entry of $\Omega_{2}$ is zero and the corresponding diagonal 
   entry of $\Omega_{1}$ differs from zero.
   Since Equation \eqref{eq:thm3} is equivalent to
   \begin{equation}
       \Pi_1 X^{\hop} \Pi_2 \Pi_2^\top \Omega_{1} \Pi_2 
       \Pi_2^\top \cop{X} \Pi_1^ \top \text{ is diagonal,}
   \end{equation}
   and
   \begin{equation}
       \Pi_1 X^{\hop} \Pi_2 \Pi_2^\top  \Omega_{2} \Pi_2 
       \Pi_2^\top X \Pi_1^\top \text{ is diagonal},
   \end{equation}
   for any permutation matrices $\Pi_1, \Pi_2$, we assume without 
   loss of generality that
   \begin{equation} 
   \label{eq:structure}
       X = \left[ \begin{array}{c|c} \widetilde{X} & \vec{x}_{1} 
       \\ \hline \vec{x}_{2}^\hop & x
       \end{array} \right],
       \Omega_{1} = \left[ \begin{array}{c|c} \widetilde{\Omega}_{1} 
       & 0 \\\hline 0 & \omega_m \end{array} \right],
       \Omega_{2} = \left[ \begin{array}{c|c} \widetilde{\Omega}_{2} 
       & 0 \\ \hline 0 & 0 \end{array} \right],
   \end{equation}
   where $\omega_m \neq 0$ and $\widetilde{X},\widetilde{\Omega}_{2}
   \in Gl(m-1)$.
   Now
   \begin{equation}
       X^{\hop} \Omega_{2} X = \left[ \begin{array}{c|c} 
       \widetilde{X}^{\hop} 
       \widetilde{\Omega}_{2}\widetilde{X} & \widetilde{X}^{\hop}
       \widetilde{\Omega}_{2} \vec{x}_{1}
       \\ \hline  \star & \star \end{array} \right],
   \end{equation}
   and Equation~\eqref{eq:thm3} together with the invertibility 
   assumption on $\widetilde{X}$ and 
   $\widetilde{\Omega}_{2}$ implies $\vec{x}_{1}=0$ and $x \neq 0$. Thus
   \begin{equation}
		X^{\hop} \Omega_{1} X^{*} = 
		\left[ \begin{array}{c|c}
		\widetilde{X}^{\hop} \widetilde{\Omega}_{1} \widetilde{X}^{*}
		+ \omega_m \vec{x}_{2} \vec{x}_{2}^{\top} & 
		x \omega_{m} \vec{x}_{2} 
		\\ \hline  \star & \star
		\end{array} \right],
   \end{equation}
   and since $x \neq 0$ and $\omega_{m} \neq 0$, Equation \eqref{eq:thm3} 
   yields that $\vec{x}_{2}=0$.
   Hence, we just showed that if $\Omega_{1}$ and $\Omega_{2}$
   are structured as in Equation \eqref{eq:structure}, 
   $X$ can only be a common diagonalizer if
   \begin{equation}
       X = \left[ \begin{array}{c|c} \widetilde{X} & 0 \\ \hline 0 & x
       \end{array} \right].
   \end{equation}
   Now, it is clear that $X \in Gl(m)\setminus\mathcal{G}(m)$ if and only if 
   $\widetilde{X} \in Gl(m-1)\setminus\mathcal{G}(m-1)$, and we reduced 
   the problem to the invertible case treated above.
   This concludes the proof of the theorem.

\subsection{Proof of Theorem \ref{thm:simult_diagpart2}}

   Again, we firstly consider the case $m=2$. Assumption 
   \eqref{eq:cond_thm3} is equivalent to Condition \eqref{eq:51a} and due 
   to the fact that Equations~\eqref{eq:proofeq2} and \eqref{eq:proofeq5} have 
   both nontrivial solutions, say
   \begin{equation}
       [x_{1} x_{2}, x_{3} x_{4}]^{\top} \text{ and } 
       [x_{1} x_{2}^{*}, x_{3} x_{4}^{*}]^\top.
   \end{equation}
   Thus, we have
   \begin{equation}
   \label{eq:profthm3001}
       \left\{\!\!
       \begin{array}{ll}
           x_{1} x_{2} \vec{\omega}_{1} + x_{3} x_{4} \vec{\omega}_2 \!\!\!\!&=0, 
           \\
           x_{1} x_{2}^{*} \vec{\omega}_1 + x_{3} x_{4}^{*} 
           \vec{\omega}_2 \!\!\!\!&=0 
       \end{array}
       \right.
   \end{equation}
   and, by taking absolute values,
   \begin{equation}
       \left\{\!\!
       \begin{array}{ll}
           | x_{1} x_{2}| \| \vec{\omega}_1 \| \!\!\!\!&= 
           | x_{3} x_{4} | \|\vec{\omega}_2\|, \\
           | x_{1} x_{2}^{*} |  \| \vec{\omega}_1 \| \!\!\!\!&=
           | x_{3} x_{4}^{*} | \| \vec{\omega}_2\| 
       \end{array}
       \right.
   \end{equation}
   and Condition~\eqref{eq:51b} follows. 

   To see the reverse direction, let Condition~\eqref{eq:51b} hold true. 
   If $\|\vec{\omega}_1\|= \|\vec{\omega}_2 \|=0$ or
   if $\|\vec{\omega}'_1\|= \|\vec{\omega}'_2 \|=0$, 
   the non-uniqueness of $X$ follows from Theorem~\ref{thm:01}. 
   Otherwise, (ii) implies (after a possible renumeration)
   \begin{equation}
       \begin{split}
       \vec{\omega}_{2}  & = r {\rm e}^{{\rm i} \varphi_1} \vec{\omega}_{1} \\
       \vec{\omega}'_{2} & = r \vec{\omega}'_{1},
   \end{split}
   \end{equation}
   with $r>0$ and $\varphi_1 \in [0,2\pi)$.
   Using Equation \eqref{eq:profthm3001}, we find an explicit diagonalizer 
   that is not in $\mathcal{G}(m)$, namely
   \begin{equation}
   \begin{split}
   & x_1=\widetilde{r} {\rm exp}(\tfrac{\rm i}{2}\varphi_1), \quad
   x_2=\textstyle\frac{1}{r}, \\
   & x_3=\textstyle\frac{1}{\widetilde{r}} {\rm exp}(\tfrac{\rm i}{2}\varphi_1), \quad
   x_4=-1,
   \end{split}
   \end{equation}
   where $\widetilde{r}\neq 0$ can be chosen arbitrarily such that 
   $X$ is invertible.

   Let us consider now the case $m>2$.
   If there exists a pair $(k,l)$ with $k\neq l$, such that 
   Conditions~\eqref{eq:conditionlast} hold, then 
we can use the above argument for the corresponding $(2 \times 2)$-sub matrix
and conclude that the common diagonalizer is not essentially unique.
   
   Now, let $X \in Gl(m)\setminus \mathcal{G}(m)$. 
   Assume for the moment that at least one per $C_{i}$'s and $C'_{j}$'s is 
   invertible, say, $C_1$ and $C'_1$.
   This implies that  $C_i C'_j C_1^{-1} C_1^{'-1}= X^{\hop} \Omega_i \Omega'_j 
   \Omega_1^{-1}{\Omega}_1^{'-1}X^{-1}$, for $i=1, \ldots, s$ and $j=1, \ldots, h$,
   is a simultaneous eigendecomposition. 
   Since $X \in Gl(m)\setminus \mathcal{G}(m)$, there must be an 
   index pair $(k,l)$ with $k \neq l$, such that
   \begin{equation}
       \frac{\omega_{ik} \omega'_{jk}}{\omega_{1k}\omega'_{1k}}=\frac{\omega_{il}\omega'_{jl}}{\omega_{1l}\omega'_{1l}}
   \end{equation}
   for all $i=1, \ldots, s$ and $j=1, \ldots, h$. 
   This yields $|c(\vec{\omega}_k,\vec{\omega}_l)|=|c(\vec{\omega}'_k,\vec{\omega}'_l)|=1$ and hence Equation (i) follows.
   If none of the $C_i$ is invertible, the same argument as in Theorem 
   \ref{thm:01} applied to both sets 
   $\{C_{i}\}_{i=1}^{s}$ and $\{C'_{j}\}_{j=1}^{h}$ individually yields the 
   same conclusion as in \eqref{eq:51a}.

   Hence, by permuting $k$ and $l$ if necessary, there exist  $z_1, z_2 \in \C$ such that
   \begin{equation}
   \label{eq:rumpelbla}
       \vec{\omega}_k = z_1 \vec{\omega}_l, \text{ and } \vec{\omega}'_k = z_2 \vec{\omega}'_l.
   \end{equation}
   On the other hand, by Theorem \ref{thm:simult_diag_TWO}, we obtain
   \begin{equation}
       | \omega_{ik} | |\omega'_{jl}|= |\omega_{il}| |\omega'_{jk}|
   \end{equation}
   for all  $i=1, \ldots, s$ and $j=1, \ldots, h$, and hence $|z_1|=|z_2|$.
   Equation \eqref{eq:rumpelbla} now yields Equation~\eqref{eq:51b} and the proof is complete.
   %
   %
%

\subsection{Proof of Lemma \ref{lem:put}}
(i) The construction of $C_{2}$ as in Equations \eqref{eq:put1} and \eqref{eq:put2} implies %
\begin{equation}
\label{eq:14}
	A \Omega_{2} A^{\top} = U \Sigma U^{\top}.
\end{equation}
As diagonal entries of $\Sigma$ are all positive, Equation~\eqref{eq:14}
is equivalent to
\begin{equation}
	\Sigma^{-1/2} U^{\hop} A \Omega_{2} A^{\top} \cop{U} \Sigma^{-1/2} = I_{m}.
\end{equation}
By inserting $\Omega_{2} = (\Omega_{2}^{1/2})^{2}$ into the above equation,
it can be seen that $V := \Sigma^{-1/2} U^{\hop} A \Omega_{2}^{1/2}$ 
is complex orthogonal. 
Now, $A = U \Sigma^{1/2} V \Omega_{2}^{-1/2}$, and thus Equations \eqref{eq:put1} and \eqref{eq:put2} yield
\begin{equation}
\label{eq:15}
	C_{1} = A \Omega_{1} A^{\hop} =
	U \Sigma^{1/2} V \underbrace{\Omega_{2}^{-1/2} \Omega_{1} \Omega_{2}^{-\hop/2}}_
	{=: \Lambda}
	V^{\hop} \Sigma^{1/2} U^{\hop},
\end{equation}
where $\Lambda$ is diagonal.
Then, Equation~\eqref{eq:15} is equivalent to
\begin{equation}
	\Sigma^{-1/2} U^{\hop} C_{1} U \Sigma^{-1/2} = 
	V \Lambda V^{\hop}.
\end{equation}
%

(ii) It is straightforward to verify that
\begin{equation}
	X^{\hop} C_{1} X = V^{\top} \Sigma^{-1/2} U^{\hop} C_{1} 
	U \Sigma^{-1/2} \cop{V} = \Lambda,
\end{equation}
and
\begin{equation}
	X^{\hop} C_{2} \cop{X} = V^{\top} \Sigma^{-1/2} U^{\hop} C_{2} 
	\cop{U} \Sigma^{-1/2} V = I_{m}.
\end{equation}

%


\end{document}